\documentclass[preprint,showpacs,aps]{revtex4}
\usepackage{graphicx}

\begin{document}

\title{Absolute frequency measurements of the $D_2$ line
and fine-structure interval in $^{39}$K}
\author{Ayan Banerjee and Vasant Natarajan}
\email{vasant@physics.iisc.ernet.in}
\affiliation{Department of Physics, Indian Institute of
Science, Bangalore 560 012, INDIA}

\begin{abstract}
We report a value for the $D_2$-line frequency of $^{39}$K
with 0.25 ppb uncertainty. The frequency is measured using
an evacuated ring-cavity resonator whose length is
calibrated against a reference laser. The $D_2$ line
presents a problem in identifying the line center because
the closely-spaced energy levels of the excited state are
not resolved. We use computer modelling of the measured
spectrum to extract the line center and obtain a value of
391 015 578.040(75) MHz. In conjunction with our previous
measurement of the $D_1$ line, we determine the
fine-structure interval in the $4P$ state to be 1 729
997.132(90) MHz. The results represent significant
improvement over previous values.
\end{abstract}


\maketitle

\section{Introduction}
Precise measurements of atomic energy levels continue to
play an important role in the development of physics. The
energy levels of alkali atoms are particularly important
because these atoms can be laser cooled to ultra-low
temperatures for subsequent use in precision spectroscopy
experiments. For example, precise measurements of the $D_1$
line in Cs \cite{URH99}, Rb or K \cite{BDN03a}, in
conjunction with photon-recoil shift measurements in an
atom interferometer \cite{WYC93}, could yield an
independent value of the fine-structure constant $\alpha$.
In addition, a precise value of the frequency of the $D_2$
line in Cs \cite{URH00} is required for
atom-interferometric measurements of the local
gravitational acceleration \cite{PCC99}.

We have recently developed a technique for measuring the
absolute frequencies of optical transitions with sub-MHz
precision \cite{BDN03,BRD03}. The frequency is measured
using a ring-cavity resonator whose length is calibrated
against a reference laser locked to the $D_2$ line of
$^{87}$Rb. The frequency of the reference laser is known to
an accuracy of 10 kHz \cite{YSJ96}. We have already used
this technique to measure the $D_1$ lines in the alkali
atoms $^{39}$K, $^{85}$Rb, and $^{87}$Rb with 0.13 ppb
uncertainty \cite{BDN03a}, to facilitate measurements of
$\alpha$. In this paper, we apply this technique to measure
the $D_2$ line of $^{39}$K with an uncertainty of 0.25 ppb.
This represents an improvement of more than two orders of
magnitude over tabulated values \cite{NIST}. Furthermore,
in combination with our previous measurement of the $D_1$
line, we obtain the fine-structure interval in the $4P$
state of $^{39}$K with very high precision. Knowledge of
fine-structure intervals is useful in the study of atomic
collisions and relativistic calculations of atomic energy
levels. Indeed, a combination of improved methods of
calculation and increasingly precise measurements of atomic
fine and hyperfine structure (e.g.\ in Li \cite{WAC03}) can
lead to a precise value of $\alpha$ \cite{FHH80}.

The $D_2$ line in K presents a problem in determining the
line center because transitions to individual hyperfine
levels of the excited state are not resolved in
conventional saturated-absorption spectroscopy. This is
because the various hyperfine levels in the $4P_{3/2}$
state lie within 30 MHz of each other \cite{AIV77}, while
the natural linewidth is 6 MHz. We have recently
demonstrated a technique to resolve such closely-spaced
transitions \cite{BAN03}, however, in this work we use
computer simulation of the measured spectrum to extract the
line center with an accuracy of $\sim$100 kHz.

\section{Experimental details}
\begin{figure}
\includegraphics[width=1\textwidth]{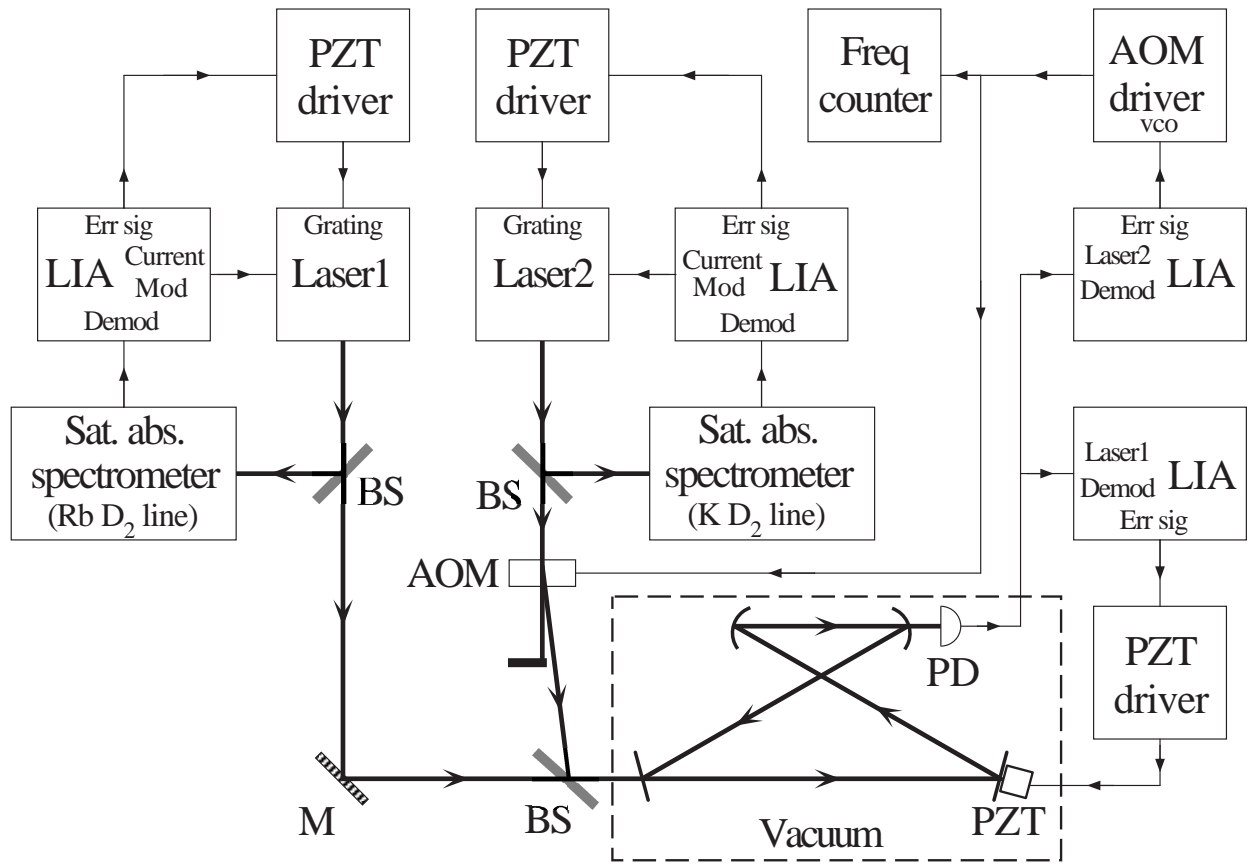}
\caption{ Schematic of the experiment. PZT, piezoelectric
transducer; AOM, acousto-optic modulator; vco,
voltage-controlled oscillator; LIA, lock-in amplifier;
Sat.\ abs., saturated absorption; BS, beam splitter; PD,
photodiode; M, mirror.} \label{f1}
\end{figure}

The experimental schematic is shown in Fig.\ \ref{f1} and
has been described extensively in a previous publication
\cite{BDN03a}. The four-mirror ring-cavity is placed inside
a vacuum chamber and evacuated to a pressure of
$\sim$10$^{-2}$ torr to eliminate frequency shifts due to
dispersion of air. Lasers 1 and 2 are standard
external-cavity diode lasers stabilized using optical
feedback from a piezo-mounted grating \cite{BRW01}. The
reference laser (Laser1) is locked to the $D_2$ line of
$^{87}$Rb using saturated-absorption spectroscopy in a
vapor cell, and the cavity is locked to the reference
laser. In general, the frequency of the laser to be
measured (Laser2) will be offset from the nearest cavity
resonance. This offset is compensated by using an
acousto-optic modulator (AOM) between Laser2 and the
cavity. The AOM is locked to this frequency difference and
its frequency is read using a counter. The exact mode
number of the cavity is determined by measuring the
cavity's free-spectral range using different lock points of
the reference laser, as described in Ref.\ \cite{BDN03}.
The mode number and the AOM offset together yield the
frequency of Laser2. A second AOM kept between the
reference laser and its saturated-absorption spectrometer
allows us to vary the reference frequency continuously so
that the AOM offset measured for Laser2 is always close to
a given value. This eliminates potential systematic errors
due to changes in the direction of the beam entering the
cavity.

The error signals needed for locking the diode lasers are
produced by modulating the injection current at a frequency
of 20--50 kHz. The error signal is obtained from the
saturated-absorption signal by phase-sensitive detection at
the third harmonic of the modulation frequency
\cite{WAL72}. This produces narrow dispersive signals that
are free from effects due to the underlying Doppler profile
or intensity fluctuations. The error signal for locking the
cavity is only a first-derivative signal since the cavity
modes appear against a flat background.

The spectroscopy on $^{39}$K is performed in an ultrahigh
vacuum glass cell maintained at a pressure below $10^{-8}$
torr by an ion pump. K vapor is produced by heating a
getter source \cite{SAES} with a current of 2.6~A. The
ultrahigh vacuum environment is necessary to minimize
linewidth broadening arising as a result of background
collisions. The getter source also gives us control over
the amount of K vapor in the cell, which we optimized to
obtain the narrowest linewidth.

\begin{figure}
\includegraphics[width=1\textwidth]{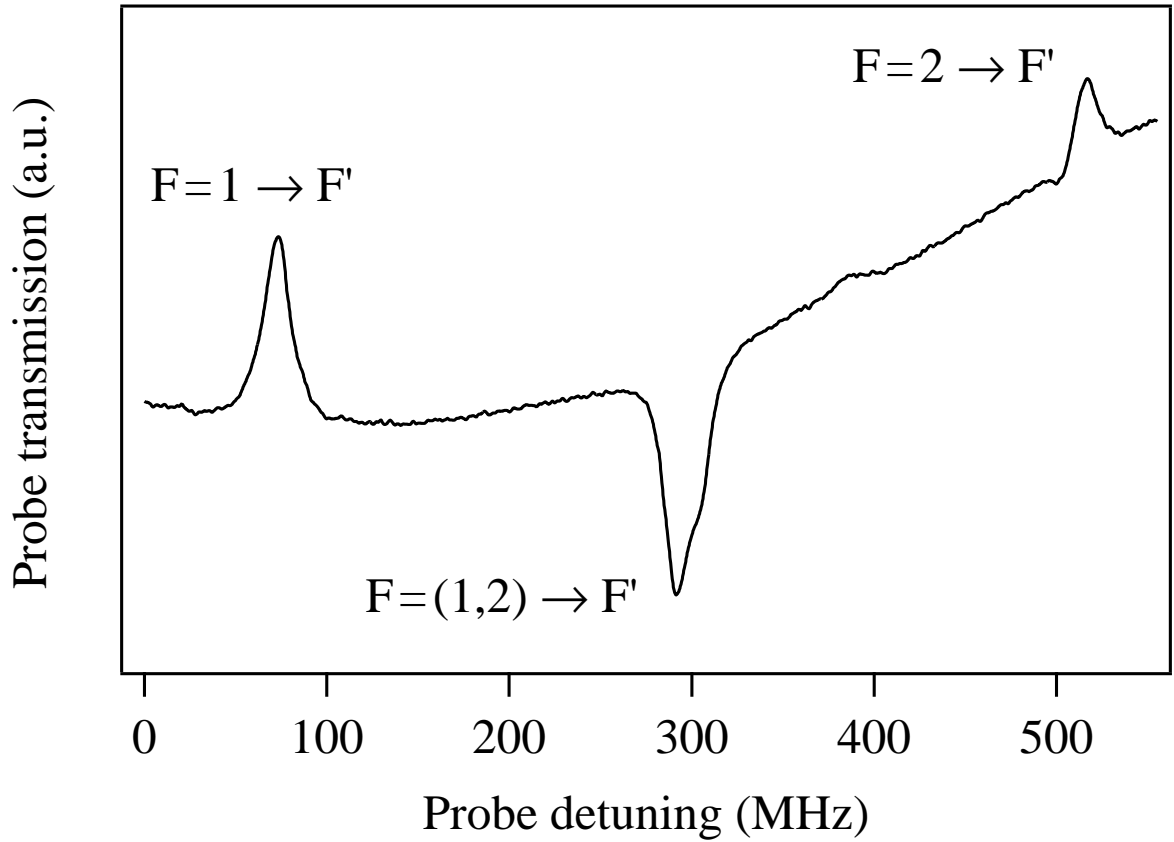}
\caption{Typical saturated-absorption spectrum of the $D_2$
line in $^{39}$K showing all three sets of transitions. The
inverted peak in the center is a ground crossover
resonance.} \label{f2}
\end{figure}

Fig.\ \ref{f2} shows a typical saturated-absorption
spectrum of the $D_2$ line covering the two
ground-hyperfine levels. Each peak is actually a
convolution of six peaks. However, the individual hyperfine
transitions are not resolved because the different
hyperfine levels of the excited state lie within 30 MHz of
each other. To determine the line center, we therefore
performed computer simulations of the spectrum. We first
fixed the locations of the 6 peaks (with respect to the
line center) according to the known hyperfine shifts
\cite{AIV77}. We then set the peak amplitudes to correspond
to those obtained in saturated-absorption spectroscopy. We
further assumed that the linewidth of all the transitions
is the same. We calculated the spectrum for a given value
of the line center and linewidth.

\begin{figure}
\includegraphics[width=1\textwidth]{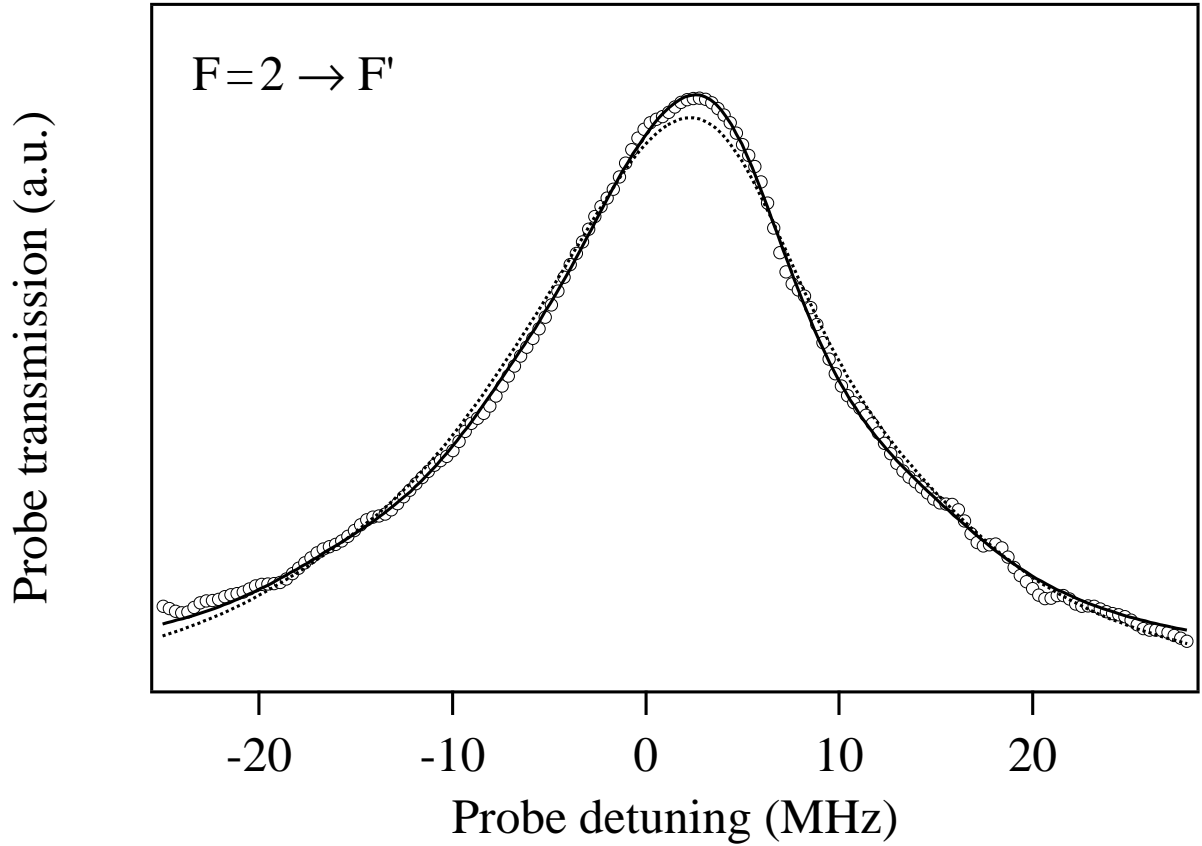}
\caption{Measured and calculated spectra for $F=2
\rightarrow F'$ transitions. The measured spectrum, shown
as open circles, is a convolution of 6 peaks. The solid
curve is the best fit obtained with a linewidth of 12.3
MHz. The dotted curve is the calculated spectrum with a
linewidth of 15 MHz. Even with such a small change in
linewidth, the curve does not fit the measured spectrum
very well.} \label{f3}
\end{figure}

Fig.\ \ref{f3} shows a close-up of the measured spectrum
(open circles) for $F=2 \rightarrow F'$ transitions. The
solid curve is a best fit to the spectrum, obtained with a
linewidth of 12.3 MHz. In order to check that this
linewidth is reasonable, we tuned the diode laser to the
$D_1$ line of K (at 770 nm), where the individual hyperfine
transitions are clearly resolved. The linewidth obtained
for those transitions was 14 MHz, close to the fit value
for the $D_2$ line. To check the linewidth further, we
changed the linewidth for the calculated spectrum and held
it constant during the fitting. The best-fit calculated
spectra for a linewidth of 15 MHz is also shown in Fig.\ 3.
Even with this small change, the lineshape deviates from
the measured data. The fit linewidth of 12.3 MHz, however,
is larger than the natural linewidth of 6 MHz. The primary
causes for this increase are power broadening due to the
pump beam and a small angle between the counterpropagating
pump and probe beams. The effect of stray magnetic fields
and background collisions is negligible.

With the best fit to the spectrum in Fig.\ 3, we found that
its maximum lies 2.60(10) MHz above the line center.
Similar modelling for transitions starting from the $F=1$
ground level shows that the maximum in the spectrum lies
12.03(10) MHz below the line center.

\section{Error analysis}
The errors in our frequency measurement technique have been
discussed extensively in earlier publications
\cite{BDN03a,BDN03,BRD03}. We present here a brief overview
for the sake of completeness. There are two classes of
potential systematic errors that we consider. The first
class of errors comes from systematic shifts in the laser
frequencies. For the reference laser, changes in the
lineshape of the peaks due to optical pumping effects are
taken care of by carefully adjusting the pump and probe
beam intensities in the saturated-absorption spectrometer
(to a ratio of about 3). Shift in its lock point due to
peak pulling from neighboring transitions, the underlying
Doppler profile, or phase shifts in the feedback loop are
minimized by third-harmonic detection for the error signal.
Collisional shifts in the Rb vapor cell and the effect of
stray magnetic fields are negligible.

The second class of systematic errors is inherent to our
technique because we are really comparing the wavelength
(and not the frequency) of the two lasers. The most
important source of error is dispersion inside the cavity,
which is eliminated by using an evacuated cavity. However,
there could be wavelength-dependent phase shifts at the
dielectric coated mirrors used in the cavity. Such errors
can be corrected by repeating the measurement at different
cavity lengths. We have shown earlier \cite{BDN03a} that
this error is negligible when the unknown laser differs
from the reference laser by up to 25 nm. In the current
work, the wavelength difference was only 13 nm, therefore
the measurements were done at a single cavity length of
$\sim$178 mm.

\section{Results}
The results of our measurements are listed in Table I. Each
value is an average of 50--100 individual measurements, and
the quoted error includes our estimate of the systematic
error. To check for long-term variations, we repeated these
measurements over a period of several weeks. In addition,
we used two different lock points of the reference laser,
namely the $F=2 \rightarrow F'=(2,3)$ transition and the
$F=1 \rightarrow F'=(1,2)$ transition. The frequencies for
these transitions differ by 6622.886 MHz. The consistency
of our values for the two cases acts as a check on our
ability to determine the correct cavity mode number since
the cavity free-spectral range is only about 1.7 GHz.
Another check on our results is that the measured
frequencies for the two transitions should differ by the
ground hyperfine splitting in $^{39}$K. Our value of
461.569(210) MHz overlaps very well with the accepted value
of 461.720 MHz \cite{AIV77}.

The values in Table I can be combined to yield the
hyperfine-free frequency of the $D_2$ line in K
\begin{center}
$4P_{3/2} - 4S_{1/2}$ : \ \ 391 015 578.040(75) MHz
\end{center}
This value is consistent with the value of 391 015 585(30)
MHz from the National Institute of Standards and Technology
energy-level tables \cite{NIST}, but the accuracy is
improved by more than two orders of magnitude. The above
result can be combined with our earlier measurement of the
$D_1$ line \cite{BDN03a} to yield the fine-structure
interval in the $D$ line in K
\begin{center}
$4P_{3/2} - 4P_{1/2}$ : \ \ 1 729 997.132(90) MHz
\end{center}
This value overlaps at the $1\sigma$ level with our earlier
measurement of the fine-structure interval using a
home-built wavemeter \cite{BDN04}: 1 729 993 (5) MHz, but
again the accuracy is improved considerably.

\section{Conclusion}
In conclusion, we have measured the $D_2$-line frequency
and fine-structure interval in $^{39}{\rm K}$ with
uncertainty less than 100 kHz. The measurement is
complicated by the fact that the closely-spaced hyperfine
transitions in the line are not resolved in conventional
saturated-absorption spectroscopy. We use computer
simulation of the measured spectrum to extract the line
center. The frequency is measured using a Rb-stabilized
ring-cavity resonator that has general applicability to
precise frequency measurement of optical transitions
\cite{BDN03}. The only other technique of measuring optical
frequencies with comparable precision is the frequency-comb
method \cite{URH99}. By comparison, our technique is
low-cost and easier to implement. In the future, we plan to
measure the $D$ lines and fine-structure interval of other
alkali atoms such as Li, Na, and Cs. These measurements
could play an important role in QED-independent
determinations of the fine-structure constant $\alpha$.

This work was supported by the Board of Research in Nuclear
Sciences (DAE), and the Department of Science and Technology,
Government of India.

\newpage

\begin{table}
\caption{Measured frequencies for the center of different
transitions in the $D_2$ line of $^{39}$K. The frequencies
were measured with two lock points of the reference laser:
Ref.\ $f_1$ denotes the $F=2 \rightarrow F'=(2,3)$
transition and Ref.\ $f_2$ denotes the $F=1 \rightarrow
F'=(1,2)$ transition. \label{t1}}
\begin{tabular}{crr}
\hline\hline
\multicolumn{1}{c}{\parbox[t]{2.2cm}{Measured \\
transition}} &
\multicolumn{1}{c}{\parbox[t]{2.5cm}{Frequency (MHz) Ref.\
$f_1$}} & \multicolumn{1}{c}{\parbox[t]{2.5cm}{Frequency
(MHz) Ref.\ $f_2$}} \\
\hline $ F=2 \rightarrow F'$ & 391 015 404.938(150) & 391
015 404.994(150) \\
 $ F=1 \rightarrow F'$ & 391 015 866.574(150) &
391 015 866.495(150) \\
\hline\hline
\end{tabular}
\end{table}

\vspace*{6cm}

\newpage

\end{document}